# Cooperative Inference with Interleaved Operator Partitioning for CNNs

Zhibang Liu[1], Chaonong Xu[1(✉)], Zhizhuo Liu[1], Lekai Huang[1],
Jiachen Wei[1] and Chao Li[2]

[1] Beijing Key Lab of Petroleum Data Mining,
China University of Petroleum, Beijing, 102249, China
xuchaonong@cup.edu.cn
[2] Research Center for Space Computing System,
Zhejiang Lab, Hangzhou, 311121, China

**Abstract.** Deploying deep learning models on Internet of Things (IoT) devices often faces challenges due to limited memory resources and computing capabilities. Cooperative inference is an important method for addressing this issue, requiring the partitioning and distributive deployment of an intelligent model. To perform horizontal partitions, existing cooperative inference methods take either the output channel of operators or the height and width of feature maps as the partition dimensions. In this manner, since the activation of operators is distributed, they have to be concatenated together before being fed to the next operator, which incurs the delay for cooperative inference. In this paper, we propose the Interleaved Operator Partitioning (IOP) strategy for CNN models. By partitioning an operator based on the output channel dimension and its successive operator based on the input channel dimension, activation concatenation becomes unnecessary, thereby reducing the number of communication connections, which consequently reduces cooperative inference delay. Based on IOP, we further present a model segmentation algorithm for minimizing cooperative inference time, which greedily selects operators for IOP pairing based on the inference delay benefit harvested. Experimental results demonstrate that compared with the state-of-the-art partition approaches used in CoEdge, the IOP strategy achieves 6.39% ~ 16.83% faster acceleration and reduces peak memory footprint by 21.22% ~ 49.98% for three classical image classification models.

## 1 Introduction

Artificial Internet of Things (AIoT) has been widely applied in various fields, including industrial production, autonomous driving, smart home appliances, and other miscellaneous domains [1]. With the rise of deep learning technology, the computational and memory requirements of these models have steadily increased [2]. Meanwhile, deployment and real-time inference of intelligent models have received increasing attention for AIoT [3]. On the one hand, the memory budget of IoT devices is usually limited [4], which hinders model deployments. On the other hand, many practical application scenarios demand strict real-time responsiveness for model inference, such as valve leakage detection, which may require response times at the millisecond level. If the result is too late, it can cause serious safety hazards [5].

Cooperative computing for distributed inference, i.e., cooperative inference, is an effective solution [6]. As shown in Fig. 1(a), all operators of the model will be inference sequentially on one device for the centralized inference scenario. Meanwhile, in the horizontal cooperative inference scenario depicted in Fig. 1(b), the operator of Convolution 1 is partitioned horizontally into three parts across the output channel (OC) dimension and assigned to devices A, B, and C, respectively. They execute different parts of Convolution 1 in parallel. After the execution, they broadcast and concatenate the three parts of the output, which will fed into the next operator, i.e., Convolution 2. Similar processing for Convolution 2 and Convolution 3 can also be executed. In this manner, both the overall inference time and the memory footprint will be reduced due to the parallel execution.

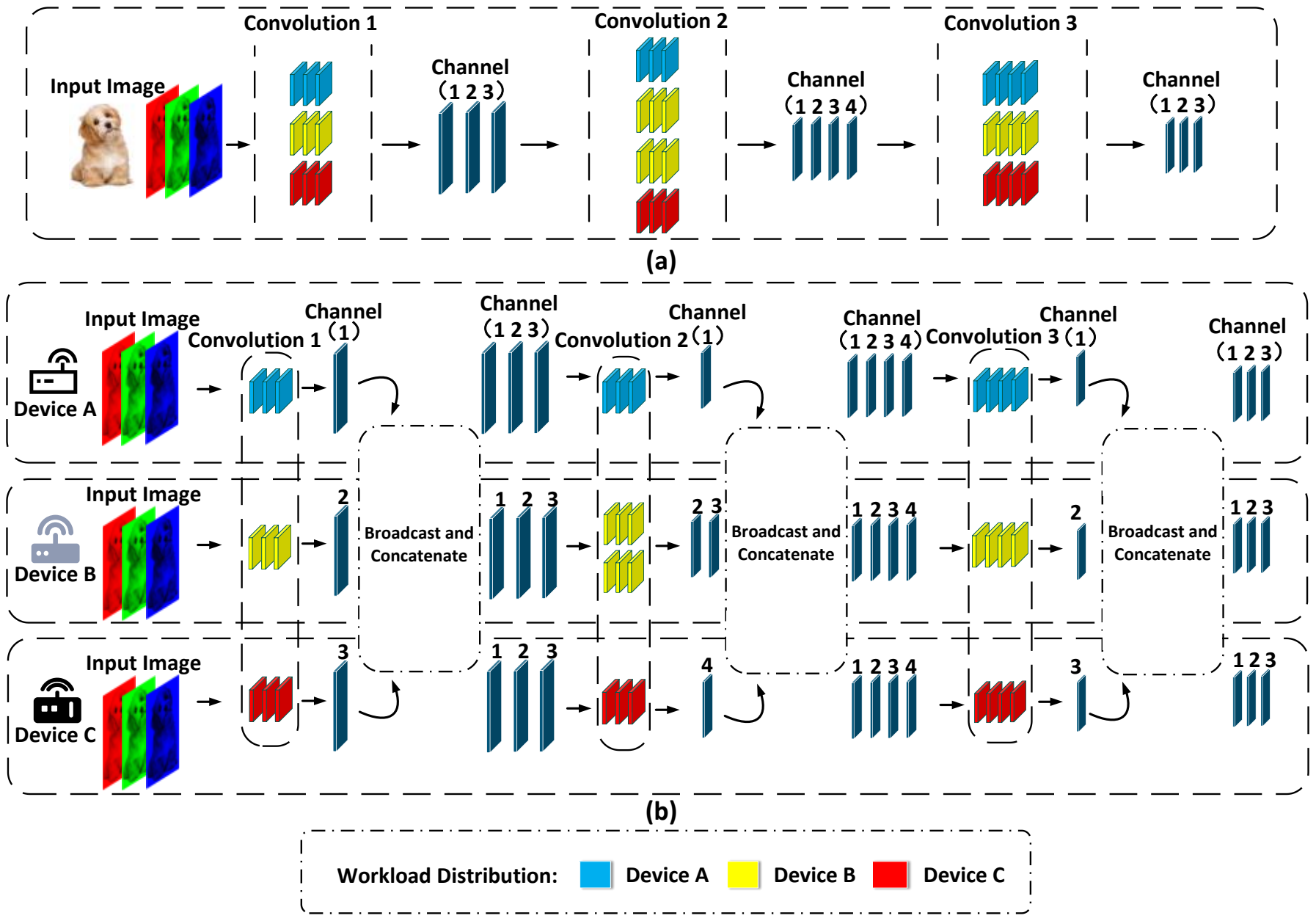

**Fig. 1.** (a) Centralized inference. (b) Cooperative inference workflow of OC dimension partitioning. After each operator execution, devices A, B, and C broadcast and concatenate the three parts of the output.

However, for an operator partitioned in cooperative inference, its output activations are distributed across multiple AIoT devices. The inference process for the next operator can only proceed once these distributed activations are concatenated [7]. This concatenation generates additional communication overhead, thereby increasing inference latency.

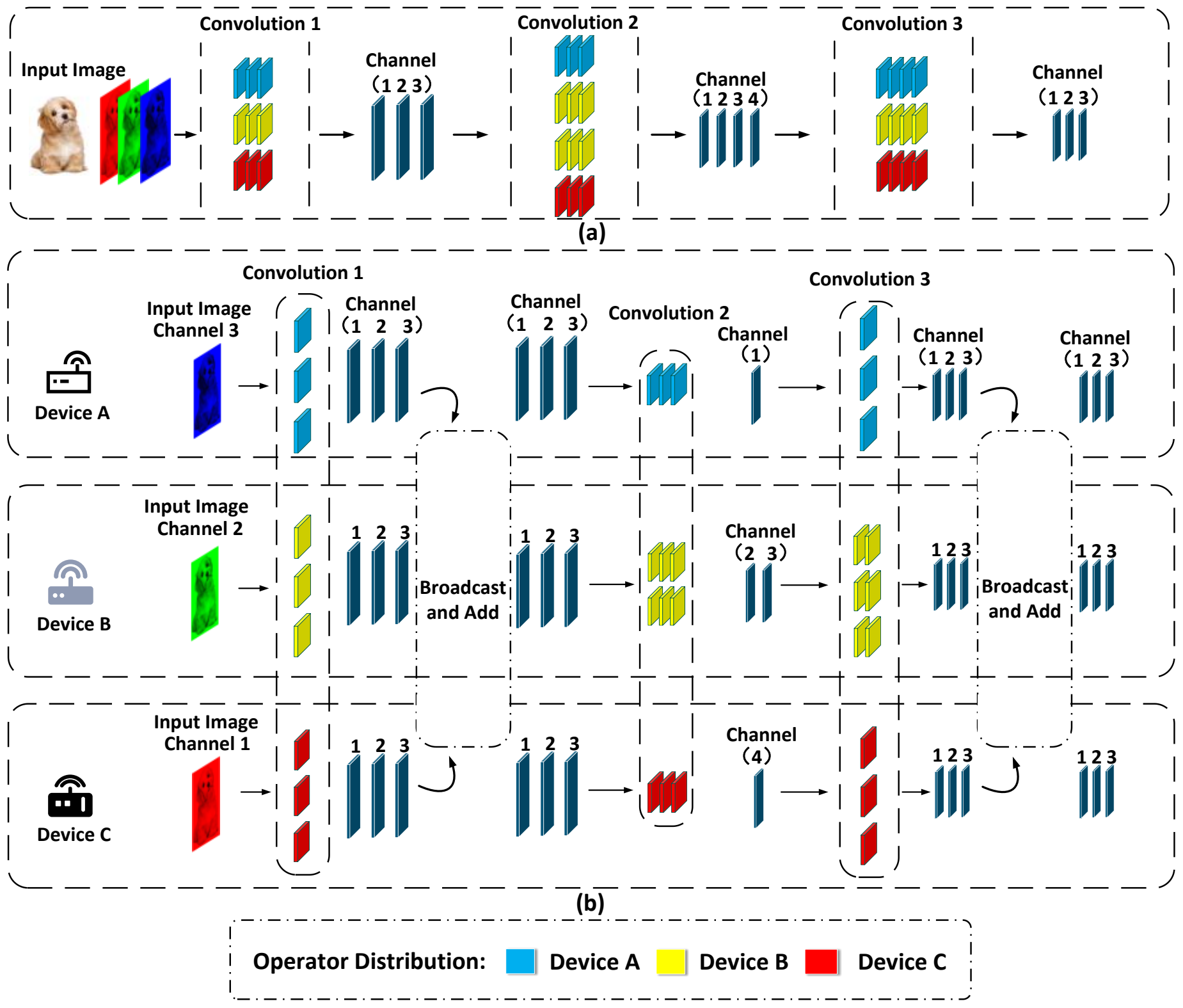

**Fig. 2.** (a) Centralized inference. (b) Cooperative CNN inference workflow of IOP. The output from Convolution 2, which is distributed on devices A, B, and C, can be directly fed into Convolution 3.

In this paper, we propose Interleaved Operator Partitioning (IOP), a novel cooperative inference scheme for AIoT. Unlike the traditional horizontal cooperative inference scheme based on the OC dimension, by partitioning an operator based on the input channel (IC) dimension and its successive operator based on the OC dimension, activation concatenation is unnecessary. Thus, cooperative inference delay is decreased. Take two adjacent operators in Fig. 2, for example. If Convolution 2 is partitioned based on the OC dimension, and Convolution 3 is on the IC dimension, the output results from Convolution 2 do not need to be concatenated before executing Convolution 3. Instead, the output results of Convolution 2 can be directly fed into Convolution 3 on the same device, thus saving one communication process and reducing the communication overhead.

In summary, this paper makes the following contributions:

- We propose IOP, a universal cooperative inference acceleration method for CNN in AIoT, which reduces inference time by minimizing the number of communications required during inference.
- We formulate the problem of minimizing inference delay based on the IOP scheme.

• We propose a heuristic model segmentation algorithm, which applies IOP across all segments comprising two operators to produce minimal cooperative inference delay.
• We evaluate the IOP strategy using multiple CNN models to demonstrate its superior performance.

The rest of this paper is organized as follows: Section 2 briefly describes the related work on cooperative inference. Section 3 provides a problem formulation using the IOP scheme. Section 4 designs the heuristic model segmentation algorithm. Section 5 conducts an experimental evaluation and analysis, and Section 6 is the conclusion.

## 2 Related Work

Cooperative inference involves the partitioning and allocation of the DNN computational workload while minimizing the communication overhead generated due to the distribution of computational workload [8]. Three kinds of partitioning methods are usually adopted including feature map partitioning, model partitioning and operator partitioning. Typical feature map partition strategies are used in MoDNN [9], CoEdge [10], and Musical Chair [11], while model partition strategy is used in Gpipe and hybrid parallelism [12]. Meanwhile, AlexNet [13] utilizes operator partition strategy.

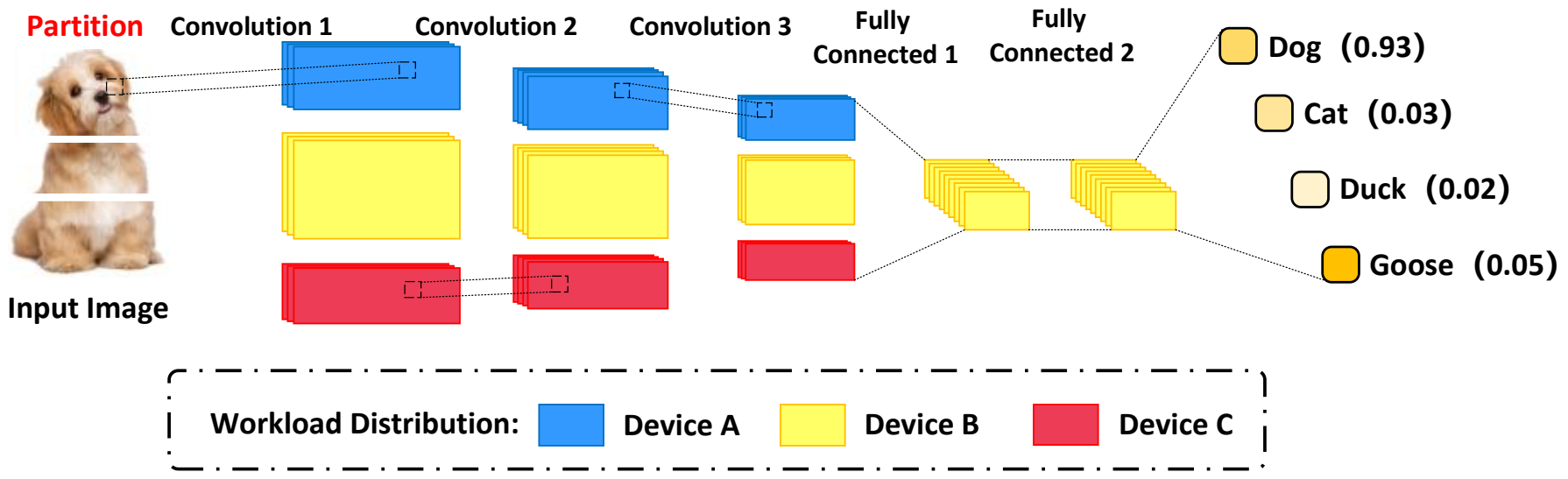

**Fig. 3.** Cooperative CNN inference workflow of CoEdge. The input image is piece-wise partitioned to patches before execution. During the convolution stage, these patches are distributed to devices A, B, and C for processing. In the fully connected stage, the activations are concatenated to complete the inference.

Feature map partitioning aims to achieve a rational distribution of workloads. For the convolution operator, MoDNN partitions the input tensor along either its rows or columns based on the device's computation capability to reduce the communication overhead. CoEdge reduces communication overhead by setting a minimum for the number of rows partitioned, thus minimizing tensor padding at partition boundaries. As shown in Fig. 3, CoEdge partitions the activations among the devices. However, due to the sliding computational process used by convolution operators, partitioning the feature map along the height (H) or width (W) dimensions requires adjacent devices to exchange data at the partition boundaries.

Operator partition commonly uses the OC in the partitioning dimension. To reduce inference latency and device memory usage, Alex Krizhevsky partitioned the model's operators across two devices along the OC dimension. However, this method requires distributed activations to be broadcast and concatenated before being fed to the next operator, resulting in significant communication overhead.

To tackle this challenge, we propose an IOP scheme that partitions an operator based on the IC dimension and the subsequent operator based on the OC dimension. In this case, activation concatenation becomes unnecessary, which reduces the number of communication links. This reduction leads to a decrease in the cooperative inference delay.

## 3 Problem Formulation

We assumed that the communication bandwidth and computing capability of all devices are relatively stable. To ensure a precise problem formulation, the following necessary concepts and symbols are defined:

$N = [1,2, \ldots, n]$ denotes the operator in order in the pre-deployed model, and the $M = [1,2, \ldots, m]$ denotes the set of available devices.

$(f, r)_j$ is employed to describe the information of computing devices $j$, where $f$ represents the device's computing capability, and $r$ represents its available memory capacity. The communication bandwidth between devices is represented by $b$.

$\eta_i = (H, IC, OC)_i$ is employed to describe partition dimension, where $H, IC, OC \in \{0,1\}$, represent the selection of the partition dimension of operator $o_i$, $H$ is the feature map H dimension and $IC$, $OC$ represent the IC and OC dimension of operator, respectively.

$(c^{in}, c^{out}, w^k, h^k, s, p)_i$ is employed to describe operator parameter. For a convolution operator, $c^{in}$ is the IC number, and $c^{out}$ is the OC number. $w^k$ is the convolution kernel width, $h^k$ is the kernel height, $s$ is the stride and $p$ is the padding size. Fully connected operators can be considered a special convolution operator $c^{in}$ with the input dimension and $c^{out}$ as the output dimension.

$o_{i,j}$ represents the operator is partitioned into multiple parts for the part of $o_i$ assigned to device $j$. $c_{i,j}^{in}$ denotes its IC number, $c_{i,j}^{out}$ represents its OC number, $\omega_{i,j}$ represents its memory usage of weights, $a_{i,j}$ denotes the memory usage of its activation outputs.

There are some numerical constraints on the partition dimensions and sizes:

$$\sum_{i \in N} \omega_{i,j} + \max\left(a_{i,j}\right) \leq r_j, j \in M \tag{1}$$

Equation (1) represents the constraint that the operators deployed on each device must satisfy, which is that the peak memory footprint should be less than the device's capacity.

$$\sum_{n \in 0,1,2} \eta_i[n] = 1 \tag{2}$$

Equation (2) represents the constraint that each operator in the model can only choose one partition dimension from H, IC, and OC.

$$\sum_{j \in M} \eta_i[0] \cdot h_{i,j} = \eta_i[0] \cdot h_i, i \in N, h_{i,j} \geq 0, h_{i,j} \in Z \tag{3}$$

$$\sum_{j \in M} \eta_i[1] \cdot c_{i,j}^{in} = \eta_i[1] \cdot c_i^{in}, i \in N, c_{i,j}^{in} \geq 0, c_{i,j}^{in} \in Z \tag{4}$$

$$\sum_{j \in M} \eta_i[2] \cdot c_{i,j}^{out} = \eta_i[2] \cdot c_i^{out}, i \in N, c_{i,j}^{out} \geq 0, c_{i,j}^{out} \in Z \tag{5}$$

Equations (3), (4), and (5) represent the constraint that after partitioning, the total size of all concatenated partitions along the H, IC, and OC dimensions equals the size of the respective dimension. $\eta_i[0]$, $\eta_i[1]$ and $\eta_i[2]$ take the values 0 or 1, denotes whether to choose H, IC and OC dimensions of the operator for partitioning, respectively.

$$P1: min \sum_{\eta_i, i \in N} max_{j \in M} \left( T_{i,j}^c + T_{i,j}^g \right) \tag{6}$$

$$s.t.\ (1), (2), (3), (4), (5)$$

The total time required for model inference is comprised of two components: computation delay and communication delay, where $T_{i,j}^c$ and $T_{i,j}^g$ represent the computational and communication delay of the operator $i$ on device $j$, respectively. They can be expressed by Equations (7) and (8).

$$T_{i,j}^c = \frac{c_{i,j}}{f_j}, i \in N, j \in M \tag{7}$$

$$T_{i,j}^g = \frac{g_{i,j}}{b}, i \in N, j \in M \tag{8}$$

where, $c_{i,j}$ and $g_{i,j}$ represent the computation workload and communication workload, which depend on the method used for operator partition, while the values of $f$ and $b$ are determined by the properties of the devices.

## 4 Segmentation Algorithm

We describe the segmentation and pairing based on the IOP scheme, i.e., Problem $P1$, as an integer programming problem. To produce a feasible solution efficiently, we segment the model into $\Gamma$:

$$\Gamma = [\gamma_1, \gamma_2, \dots, \gamma_k] \tag{9}$$

where $\gamma_i$ is a segment that is either a single operator or a pair of operators. For a segment of a pair of operators, IOP will be employed on it so as to save communication overhead.

To find an optimal segmentation for minimizing cooperative inference delay. A heuristic segmentation algorithm is designed as follows: Starting from the first operator, it searches for interleaved operator pairs layer by layer. Specifically, for operator $o_i$ and its succeeding operator $o_{i+1}$, we compare the inference times using the IOP scheme and the partition approach in CoEdge. If the IOP scheme achieves a shorter inference time, the two operators will be paired to form a new segment; otherwise, a new segment will be formed that includes only $o_i$.

**Algorithm 1** Model Segmentation and Pairing Algorithm

**Input:**
Device set: $M = [1,2,\dots,m]$
Operator set: $N = [1,2,\dots,n]$
Operator parameter: $(c^{in}, c^{out}, w^k, h^k, s, p)_i$
Partition dimension tuple: $\eta_i = (H, IC, OC)_i$
Memory, frequency, and bandwidth: $(f, r)_j$, $b$

**Output:**
Pairing scheme: $\Gamma = [\gamma_1, \gamma_2, \dots, \gamma_k]$

```
Initialize i ← i + 2
for i ← 1 to n in N do
    γk ← (oi, oi+1)
    Tiop ← IOP_Partition (γk)
    Tco ←CoEdge_Partition (γk)
    if Tiop ≤ Tco then
        Add γk to Γ
        i ← i + 2, k ← k + 1
    else
        i ← i + 1
    end if
end for
return Γ
```

# 5 Performance Evaluation

We compare the IOP scheme with the following similar approaches. (1) **OC**: The layer-by-layer OC dimension partitioning method is used in the experimental prototype of the classic AlexNet network, which effectively reduces the memory footprint of a device during model inference. (2) **CoEdge**: The feature map's H dimension partitioning method effectively reduces the communication volume on one device during model inference. We implement the IOP scheme in three typical CNN models: LeNet, AlexNet, and VGG11. The details of the CNN models and the datasets used are specified in Table 1.

**Table 1.** Details of the CNNs and datasets used in the evaluation

| CNN | Description | Convolutional number | Fully connected number | Dataset |
|---|---|---|---|---|
| LeNet | 7-layer CNN | 2 | 3 | MNIST [14] |
| AlexNet | 12-layer CNN | 5 | 3 | ImageNet [15] |
| VGG11 | 17-layer CNN | 8 | 3 | ImageNet |

Fig. 4 shows the comparative results of inference latency for cooperative inference based on IOP, OC, and CoEdge, and they are measured in the same experimental settings. Compared with OC, IOP saves 31.53%, 21.06%, and 12.82% inference time for three models, respectively. Relative to that of CoEdge, IOP saves 12.05%, 16.83%, and 6.39% inference time, respectively. The cooperative inference based on OC exhibits the largest latency. The reason is that the operator partitioning along the OC dimension necessitates the aggregation of output activation for each layer.

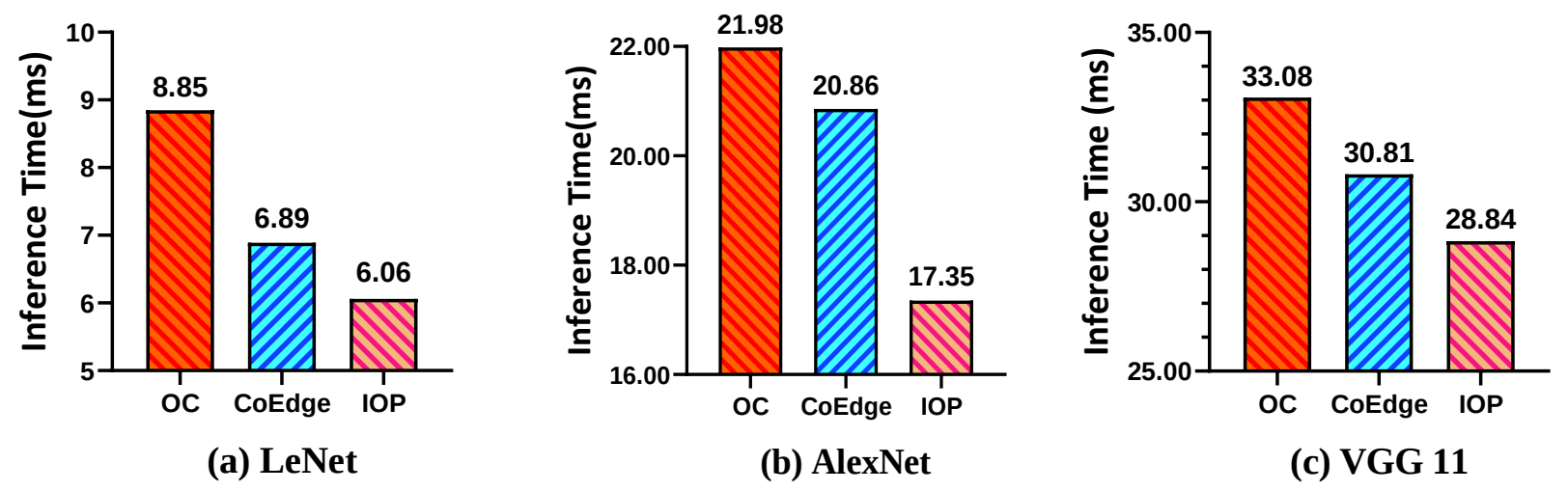

**Fig. 4.** The inference time of OC, CoEdge and IOP strategy running LeNet, AlexNet and VGG11 models.

Fig. 5 shows the peak memory footprint during model inference, where CoEdge exhibits the highest peak memory footprint since it has not partitioned the fully connected operators. In contrast, the OC and IOP schemes, which partition both convolutional and fully connected operators, exhibit lower peak memory footprints. Specifically, compared to CoEdge, IOP achieves reductions in peak memory footprint by 49.98%, 21.22%, and 40.79% for LeNet, AlexNet and VGG11, respectively.

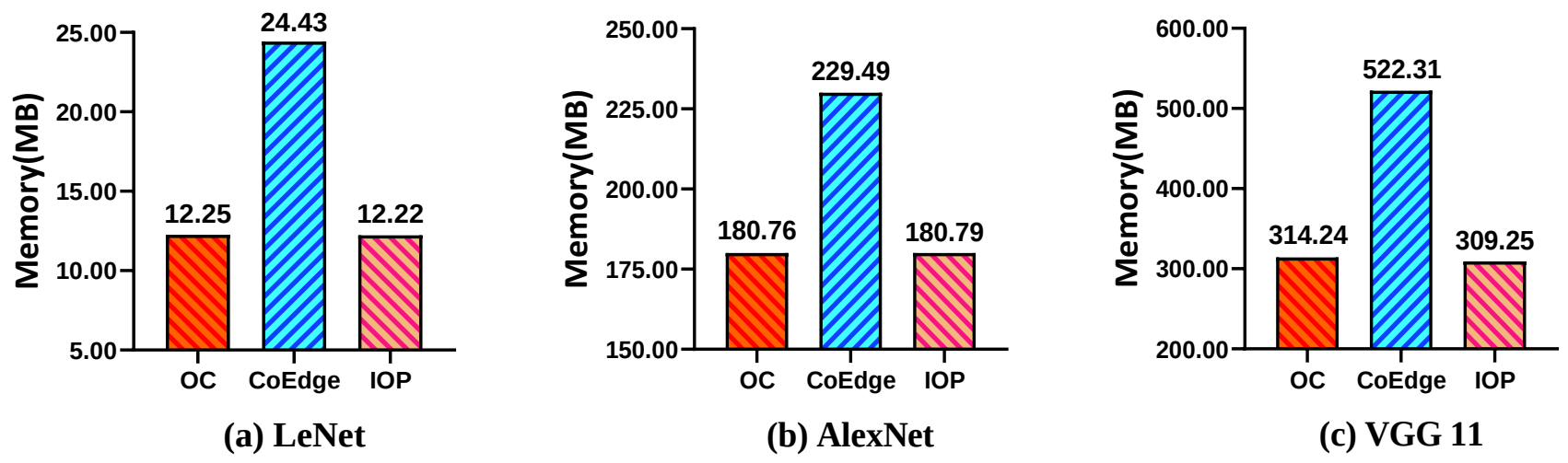

**Fig. 5.** The memory footprint of OC, CoEdge, and IOP strategy running LeNet, AlexNet and VGG11 models. All experimental settings are the same as that in the experiment Fig. 4.

We further experiments are conducted to reveal the impact of the VGG network, which is more computationally intensive relative to LeNet and AlexNet. We measure the inference time of VGG11, VGG13, VGG16, and VGG19 using OC, IOP, and CoEdge schemes with varying device communication connection establishment latency. According to the results shown in Fig. 6, inference time will increase with increasing communication connection establishment latency in all cases. For the same connection latency, IOP always achieves minimal inference time. For VGG11, the inference time using IOP is reduced by 14.51% to 26.74% when the communication connection establishment time between devices is from 1 ms to 8 ms. While for VGG13, VGG16, and VGG19, the inference time of the IOP scheme is reduced by 12.99% to 24.99%, 3.34% to 31.01%, and 15.01% to 34.87%, respectively. In other words, the larger the communication connection establishment time, the better the inference acceleration ratio of IOP relative to OC and CoEdge.

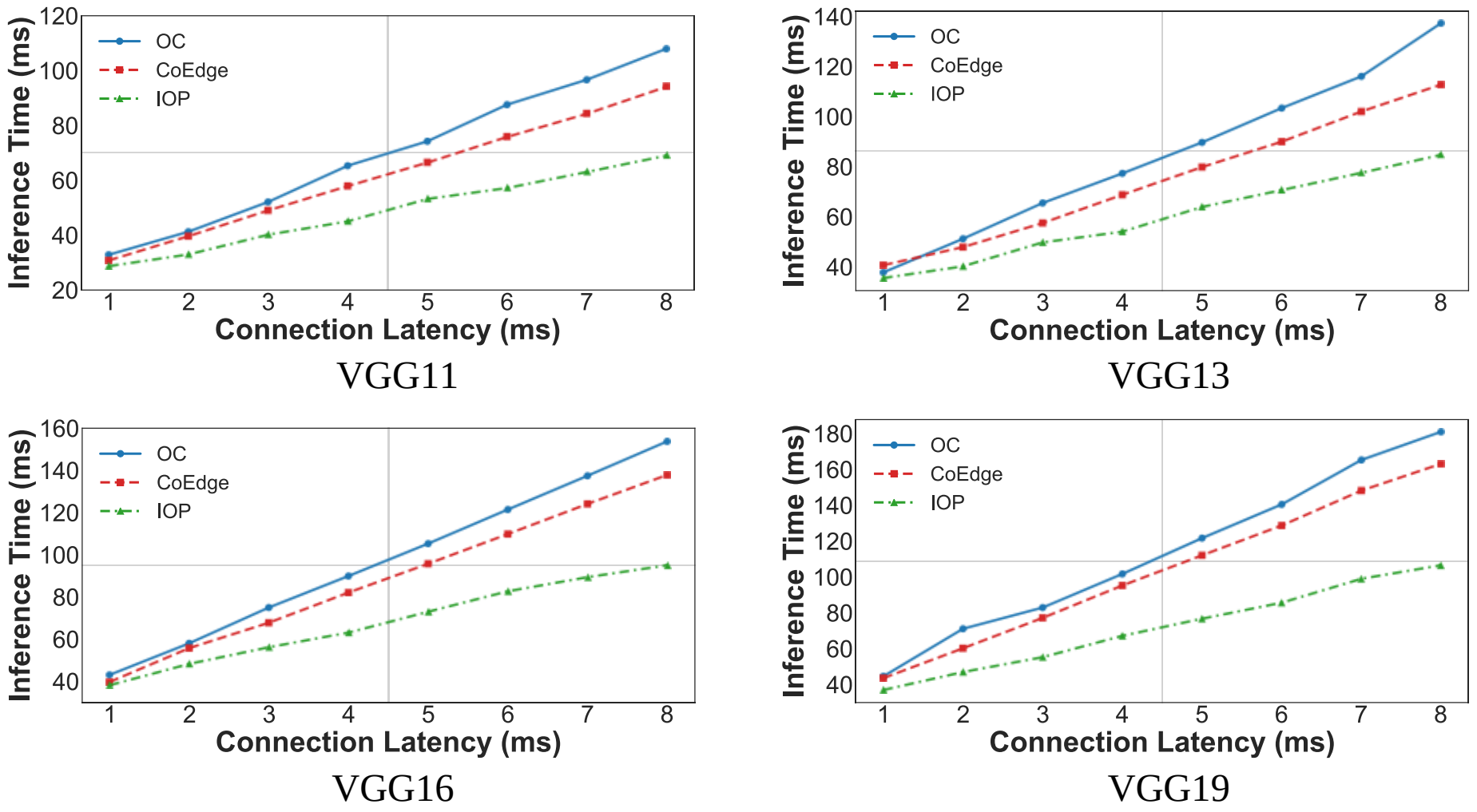

**Fig. 6.** The inference time for each of the four VGG nets under OC, CoEdge, and IOP partitioning strategies when the device communication delay ranges from 1ms to 8ms.

## 6 Conclusion

In this paper, we introduce IOP, an efficient partition strategy for cooperative inference of CNN. By employing interleaving operation partition between adjacent operators, the communication overhead caused by data sharing across devices is reduced, and thus, the inference time is decreased. We formulate the optimal distributed deployment for cooperative inference as a constrained optimization problem based on the neural network structure. To solve it efficiently, we design an operator pairing algorithm to find an efficient pairing strategy. Experimental evaluations show that for three widely adopted CNN models, including LeNet, AlexNet, and VGG11. IOP scheme achieves 6.39% ~ 16.83% inference acceleration and saves 21.22% ~ 49.98% peak memory footprint compared to the state-of-the-art CoEdge scheme.

**Acknowledgement**. This study is supported by National Key R&D Program of China (2022YFB4501600).